\documentstyle[twocolumn,prc,aps,psfig]{revtex}
\begin{document}

\title{Charmonium absorption by nucleons}
\author{A. Sibirtsev$^1$, K. Tsushima$^2$, A. W. Thomas$^2$}
\address{
$^1$Institut f\"ur Theoretische Physik, Universit\"at Giessen, \\
D-35392 Giessen, Germany \\
$^2$Special Research Center for the Subatomic Structure of Matter (CSSM) \\
and Department of Physics and Mathematical Physics, \\
University of Adelaide, SA 5005, Australia  }

\maketitle

\begin{abstract}
$J/\Psi$ dissociation in collisions with nucleons is studied within 
a boson exchange model and the energy 
dependence of the dissociation cross section is calculated 
from the threshold for $\Lambda_c{\bar D}$ production to
high energies. We illustrate the agreement
of our results with calculations based on short distance 
QCD and Regge theory. The   compatibility  between our calculations and the
data on $J/\Psi$ photoproduction on a nucleon is discussed.
We evaluate the elastic $J/\Psi{+}N$ cross section using a 
dispersion relation and demonstrate the overall agreement
with the predictions from QCD sum rules. Our results are 
compatible with the phenomenological  dissociation
cross section evaluated from the experimental data on
$J/\Psi$ production from $\gamma{+}A$, $p{+}A$ and $A{+}A$ collisions.
\end{abstract}
\vspace{2mm}

\section{Introduction}
Very recently the NA50 Collaboration reported~\cite{NA50} 
measurements of $J/\Psi$ suppression in $Pb{+}Pb$
collisions at the CERN-SPS. It was claimed in Ref.~\cite{NA50} 
that these new experimental data ruled out conventional 
hadronic models of $J/\Psi$ suppression and thus indicated
the formation of a deconfined state of quarks and gluons, 
namely the quark gluon plasma (QGP). Presently, there are 
calculations~\cite{Cassing1,Capella1,Sibirtsev9} 
using two different methods, 
namely the cascade~\cite{CassingJ,Cassing} and the Glauber type
model~\cite{Capella}, that reproduce the new 
NA50 data reasonably well up to the highest transverse
energies~\cite{NA50} based on hadronic $J/\Psi$ dissociation 
alone. In view of this, the
evidence for QGP formation based on 
anomalous $J/\Psi$ suppression~\cite{Matsui} is not obvious
and an interpretation of the  NA50 data can be 
further considered in terms of the hadronic dissociation
of the $J/\Psi$ meson by the nucleons and comovers.

However, the long term puzzle of the problem is the strength 
of the hadronic dissociation of the  $J/\Psi$ meson. The various  
calculations~\cite{Cassing1,Capella1,CassingJ,Cassing,Capella,Spieles,Kahana}
of $J/\Psi$ production from heavy ion
collisions in practice adopt different dissociation cross sections.
For instance, the cross section for $J/\Psi{+}N$ dissociation
used in these models ranges from 3.0 up to 6.7~mb when dealing 
with the same experimental data on $J/\Psi$ production
from $A{+}A$ collisions. Moreover,  it is  
presently accepted that the dissociation cross section does not
depend on the  $J/\Psi$ energy and in the available
calculations~\cite{Cassing1,Capella1,CassingJ,Cassing,Capella,Spieles,Kahana} 
it enters as a constant. Furthermore, to a certain 
extent, both of the cross sections for $J/\Psi$ dissociation 
on nucleons and comovers were taken as free parameters, finally
adjusted to the heavy ion data.

Moreover, the $J/\Psi{+}N$ cross section was evaluated 
from experimental data on $J/\Psi$ meson production in
$\gamma{+}A$ and $p{+}A$ reactions. The analysis of the  
$J/\Psi$ photoproduction from nuclei at a mean photon energy 
of 17~GeV~\cite{Anderson} indicates a $J/\Psi{+}N$ cross 
section of 3.5$\pm$0.9~mb. The combined analysis~\cite{Kharzeev4} 
of experimental data on $J/\Psi$  production from $p{+}A$ collisions 
at beam energies from 200 to 800 GeV provides a  
$J/\Psi{+}N$ cross section of 7.3$\pm$0.6 mb. On the other hand,
the $J/\Psi{+}N$ cross section evaluated by the vector dominance
model from the $\gamma{+}N{\to}J/\Psi{+}N$ data
is about 1~mb~\cite{Redlich} for $J/\Psi$ energies in 
the nucleon rest frame from 8 to 250~GeV. 

In principle, this ambiguity in the $J/\Psi$
dissociation cross section does not allow one to claim a consistent
interpretation of the NA50 results. Although the experimental
data might be well reproduced by the more recent
calculations~\cite{Cassing1,Capella1,Sibirtsev9}  
considering the  hadronic dissociation of the $J/\Psi$
meson, the $J/\Psi{+}N$ cross section was introduced
therein as a free parameter. In Ref.~\cite{Sibirtsev9} 
we provided results for $J/\Psi$ dissociation on comovers
and motivated the large rate of this process~\cite{Cassing1,Capella1} 
as arising from the in-medium modification of the dissociation
amplitude, which is obviously different from that 
given in free space~\cite{Muller,Muller1}. Most recent
results from the different models on $J/\Psi$ dissosiation
on light hadrons are given in 
Refs.~\cite{Haglin1,Haglin2,Wong,Barnes}
In the present study
we apply the hadronic model to $J/\Psi{+}N$ dissociation.

\section{Lagrangian densities, coupling constants 
and form factors}

\begin{figure}[h]
\phantom{aa}\vspace{-3mm}\hspace{-16mm}
\psfig{file=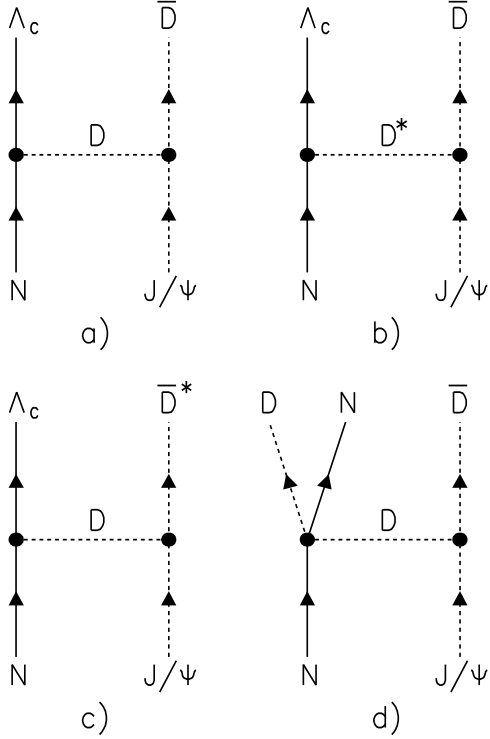,height=10.5cm,width=9.8cm}
\caption{The hadronic diagrams for $J/\Psi{+}N$ dissociation.}
\label{psich13}
\vspace{3mm}
\end{figure}

Within the boson exchange model we consider the reactions
depicted in Fig.~\ref{psich13} and use the interaction 
Lagrangian densities:

\begin{eqnarray}
{\cal L}_{J D D} &=&
i g_{J D D} \, J^\mu
\left[ \bar{D} (\partial_\mu D) - (\partial_\mu \bar{D}) D \right],
\label{pdd}\\
{\cal L}_{D N \Lambda_c} &=&
i g_{D N \Lambda_c}
\left( \bar{N} \gamma_5 \Lambda_c D
+ \bar{D} \bar{\Lambda_c} \gamma_5 N \right),
\label{dnl}\\
{\cal L}_{J D^* D} &=&
\frac{g_{J D^* D}}{m_J}\, \varepsilon_{\alpha \beta \mu \nu}
(\partial^\alpha J^\beta) \nonumber \\
& & \hspace*{3em}\times [ (\partial^\mu \bar{D}^{*\nu})\, D
                           + \bar{D} (\partial^\mu D^{*\nu}) ], 
\label{pdsd}\\
{\cal L}_{D^* N \Lambda_c} &=&
- g_{D^* N \Lambda_c}
\left( \bar{N} \gamma_\mu \Lambda_c D^{*\mu}
+ \bar{D}^{*\mu} \bar{\Lambda_c} \gamma_\mu N \right),
\label{dsnl}
\end{eqnarray}
where, $N = (\begin{array}{c} p\\ n\\ \end{array})$,
$\bar{N} = N^\dagger \gamma_0$,
$D \equiv (\begin{array}{c} D^0\\ D^+\\ \end{array})$ 
(creation of the meson states),
$\bar{D} = D^\dagger$, and similar notations for the
$D^*$ and $\bar{D}^*$ should be understood.

The $JDD$ coupling constant was derived in Refs. \cite{Muller,Ko2}
and in the following calculations we adopt $g_{JDD}$=7.64.
Following the SU(4) relations from Ref.~\cite{Ko1}
the $DN\Lambda_c$ coupling constant was taken  equal
the $KN\Lambda$ coupling constant. The 
analysis~\cite{Sibirtsev1,Sibirtsev6,MartinA,Laget,Sibirtsev7}
of experimental data provides 13.2${\le}g_{KN\Lambda}{\le}$15.7
and in the following we use $g_{DN\Lambda_c}$=14.8. However, 
we notice, that the calculations~\cite{Navarra} by the QCD sum 
rules suggest a smaller coupling $g_{DN\Lambda_c}{\simeq}6.7{\pm}2.1$.
The $D^{\ast}N\Lambda_c$ coupling constant might be again
related to the $K^{\ast}N\Lambda$ constant. The
$K^{\ast}N\Lambda$ constant is given in Ref.~\cite{Adelseck}
within the range --18.8${\le}g_{K^{\ast}N\Lambda}{\le}$--21.4, 
while it varies between $\simeq$--5.9 and --22 in Ref.~\cite{Mosel}. 
In the following calculations we adopt 
$g_{D^{\ast}N\Lambda_c}$=--19~\cite{Adelseck}.
Furthermore, we assume that $g_{JDD}{=}g_{JDD^\ast}$~\cite{Ko2}.

The form factors associated with the interaction
vertices were parameterized in a conventional monopole form
\begin{equation}
F(t)=\frac{\Lambda^2}{\Lambda^2-t}, 
\label{form}
\end{equation}
where $t$ is the four-momentum transfer and $\Lambda$ is the
cutoff parameter. The introduction of the form factors is dictated by the
extended structure of the hadrons~\cite{Machleid1}.
The form factor may be estimated through the vector dominance 
model (VDM)~\cite{Landsberg}. The VDM predicts an increase 
of the form factor with increasing mass
of the vector meson and a purely phenomenological
prediction provides $\Lambda$=$m_J$.  In the following we will
explicitly use the cutoff parameter 
$\Lambda$=3.1~GeV for the form factors at the $JDD$ and
$JDD^\ast$ vertices.   

Furthermore, one should also introduce  
form factors at the 
$DN\Lambda_c$ and $D^{\ast}N\Lambda_c$ vertices. 
Again, we use the monopole form, but there 
are no direct ways to evaluate the relevant 
cutoff parameter. In Refs.~\cite{Machleid1,Machleid2,Lohse}
the cutoff parameters were adjusted to fit the empirical
nucleon-nucleon data and range from 1.3 to 2~GeV depending on the
exchange meson coupled to the $NN$ system. Based on
these results, in the following  we use 
a monopole form factor with $\Lambda$=2~GeV for the 
$DN\Lambda_c$ and $D^{\ast}N\Lambda_c$ vertices.  

In principle, the coupling constants and the cutoff
parameters in the form factors discussed in this section 
can be adjusted to the experimental data on the
total $J/\Psi{+}N$ cross section evaluated from 
different nuclear reactions, as 
$\gamma{+}A{\to}J/\Psi{+}X$ and $p{+}A{\to}J/\Psi{+}X$.
On the other hand, these 
parameters can be also fixed by comparison to the
short range QCD or Regge theory calculations at high 
energies. As will be shown later, the set of 
coupling constants and cutoff parameters proposed 
above are rather well fitted to the relevant experimental data
and the results from theoretical calculations with
other models.

\section{The $J/\Psi{+}N{\to}\Lambda_c{+}\bar{D}$
and $J/\Psi{+}N{\to}\Lambda_c{+}\bar{D}^\ast$ reactions}

The diagrams for $J/\Psi$ dissociation by a nucleon with 
the production of the $\Lambda_c{+}{\bar D}$ 
and $\Lambda_c{+}{\bar D^\ast}$ final states are shown 
in Fig.~\ref{psich13}a)-c).
They involve the $J/\Psi{\to}D{+}{\bar D}$ and
$J/\Psi{\to}D{+}{\bar D^\ast}$
vertices and are OZI allowed. 
These reactions are endothermic, since
the total mass of the final states is larger
than the total mass of the initial $J/\Psi$-meson and nucleon.

The $J/\Psi{+}N{\to}\Lambda_c{+}\bar{D}$ reaction via $D$ 
meson exchange is shown in the Fig.~\ref{psich13}a) and
the corresponding amplitude with an amplitude added 
in order to preserve gauge invariance in the limit, 
$m_J \to 0$, may be given by:
\begin{eqnarray}
{\cal M}_a &=& 2 i g_{JDD} g_{DN\Lambda_c}
(\epsilon_J \cdot p_{\bar{D}}) \nonumber \\
& &\times \left( \frac{1}{q^2 - m_D^2} 
+ \frac{1}{2 p_J \cdot p_{\bar{D}}} \right)\,
\bar{u}_{\Lambda_c}(p_{\Lambda_c}) \gamma_5 u_N(p_N), 
\label{ma}
\end{eqnarray}
where $q = p_J - p_{\bar{D}} (= p_{\Lambda_c} - p_N)$, 
while the amplitude via the $D^\ast$ meson exchange
is shown in Fig.~\ref{psich13}b) and can be written as
\begin{eqnarray}
{\cal M}_b = 
\frac{g_{JD^*D} g_{D^*N\Lambda_c}}{m_J}\,
\frac{1}{q^2 - m_{D^*}^2} \nonumber \\
\times\, \varepsilon_{\alpha \beta \mu \nu}\,
p_J^\alpha \epsilon_J^\beta q^\mu \,
\bar{u}_{\Lambda_c}(p_{\Lambda_c}) \gamma^\nu u_N(p_N).
\label{mb}
\end{eqnarray}
Furthermore, the amplitude for the 
$J/\Psi{+}N{\to}\Lambda_c{+}\bar{D}^\ast$ reaction
due to $D$ meson exchange
is shown in Fig.~\ref{psich13}c), and is given by
\begin{eqnarray}
{\cal M}_c =
\frac{i g_{JD^*D} g_{DN\Lambda_c}}{m_J}\,
\frac{1}{q^2 - m_D^2} \nonumber\\
\times\, \varepsilon_{\alpha \beta \mu \nu}\,
p_J^\alpha \epsilon_J^\beta p_{\bar{D}^*}^\mu 
\epsilon_{\bar{D}^*}^{*\nu}\,
\bar{u}_{\Lambda_c}(p_{\Lambda_c}) \gamma_5 u_N(p_N).
\label{mc}
\end{eqnarray}
In Eqs.~(\ref{ma})-~(\ref{mc}) 
$\epsilon_J$ and $\epsilon_{\bar{D}^*}$ are respectively the polarization 
vector of the $J/\Psi$ meson and $\bar{D}^*$ meson, 
and $q{=}p_{\Lambda_c} -p_N$ 
denotes the four-momentum transfer.
We notice that there arises no interference term among the 
amplitudes, ${\cal M}_a$, ${\cal M}_b$ and ${\cal M}_c$, because of the Dirac 
structure, when spin components are not specified and summed over 
all spins for the $N$ and $\Lambda_c$, and the different final states.

In our normalization the corresponding differential cross sections 
in the center of mass frame of the $J/\Psi$
meson  and nucleon system are given  by:
\begin{eqnarray}
& &\frac{d\sigma}{d\Omega}_{a,b} = \frac{1}{64 \pi^2 s} \,
|\overline{{\cal M}_{a,b}}|^2
\nonumber \\
&\times& \left(  
\frac{[(m_{\Lambda_c}+m_D)^2 - s]
[(m_{\Lambda_c}-m_D)^2 - s]}
{[(m_N+m_J)^2 - s][(m_N-m_J)^2 - s]} \right)^{1/2} ,
\label{JNcross} \\
& &\frac{d\sigma}{d\Omega}_c
= \frac{1}{64 \pi^2 s} \, |\overline{{\cal M}_c}|^2 
\nonumber \\
&\times&  \left(
\frac{[(m_{\Lambda_c}+m_{D^*})^2 - s]
[(m_{\Lambda_c}-m_{D^*})^2 - s]}
{[(m_N+m_J)^2 - s][(m_N-m_J)^2 - s]} \right)^{1/2} ,
\label{dstar} 
\end{eqnarray}
where $s{=}(p_N{+}p_J)^2$ is the squared invariant collision
energy and 
$|\overline{{\cal M}_{a,b,c}}|^2$ are  \, the corresponding  
amplitudes \,  squared, averaged over the initial and summed over 
the final spins. Explicitly, they are given by:
\begin{eqnarray}
& & |\overline{{\cal M}_a}|^2 
= \frac{8 g_{JDD}^2 g_{DN\Lambda_c}^2}{3 m_J^2 }\, 
\left( \frac{1}{q^2 - m_D^2}
+ \frac{1}{2 p_J \cdot p_{\bar{D}}} \right)^2 \nonumber \\
&\times &
(p_N \cdot p_{\Lambda_c} - m_N m_{\Lambda_c})
[(p_J \cdot p_{\bar{D}})^2 - m_J^2 m_D^2 ] , 
\label{masquare}\\
& & |\overline{{\cal M}_b}|^2 =
\frac{g_{JD^*D}^2 g_{D^*N\Lambda_c}^2}{3 m_J^2}\,
\frac{1}{(q^2 - m_{D^*}^2)^2} \nonumber \\
&\times & 
[\,\, m_J^2 (p^2q^2-(m_{\Lambda_c}^2 - m_N^2)^2) 
\nonumber\\
& & + 2(p_J\cdot p)(p_J\cdot q)(m_{\Lambda_c}^2 - m_N^2)  
\nonumber \\
& & - p^2(p_J\cdot q)^2 -q^2(p_J\cdot p)^2
\nonumber \\
& & - 4 (p_N \cdot p_{\Lambda_c} - m_N m_{\Lambda_c})
(m_J^2q^2 - (p_J\cdot q)^2) \,\, 
] , \quad
\label{mbsquare}\\
& &|\overline{{\cal M}_c}|^2
= \frac{4 g_{JD^*D}^2 g_{DN\Lambda_c}^2}{3 m_J^2}\,
\frac{1}{(q^2 - m_D^2)^2} \nonumber \\
&\times & 
 (p_N \cdot p_{\Lambda_c} - m_N m_{\Lambda_c})
[ (p_J \cdot p_{\bar{D}^*})^2 - m_J^2 m_{D^*}^2 ],
\label{mcsquare}
\end{eqnarray}
with $p {\equiv} p_{\Lambda_c} + p_N$.

\begin{figure}[h]
\phantom{aa}\vspace{-8mm}
\psfig{file=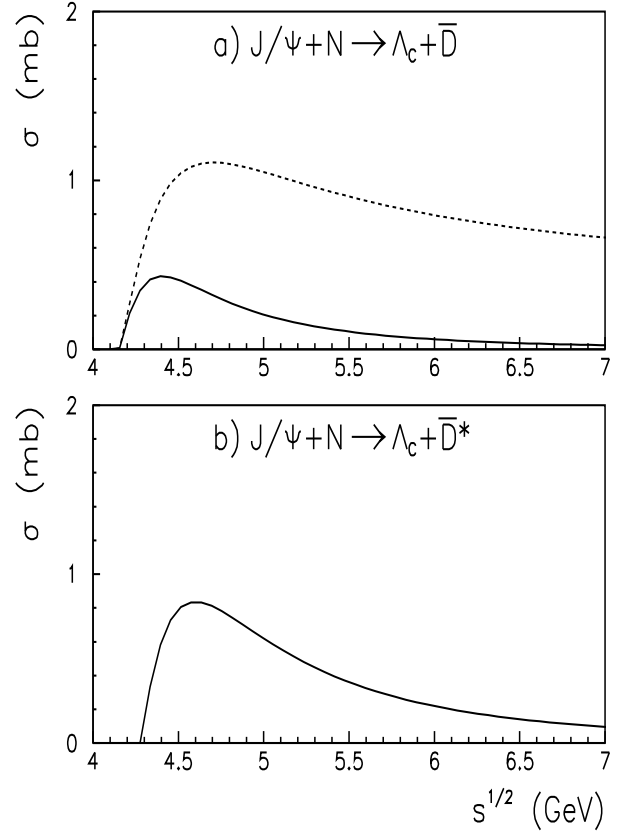,height=12.4cm,width=9.2cm}
\caption[]{The cross section for $J/\Psi$ dissociation by 
the nucleon with ${\Lambda_c{+}\bar D}$ (a)
and ${\Lambda_c{+}\bar D^\ast}$ (b) production, as a function
of invariant collision energy, $\sqrt{s}$. 
The dashed line in a) shows the calculations involving 
$D^\ast$ meson exchange, while the solid line in a) 
indicates the contribution from the $D$ exchange.
The results are shown for calculations with form 
factors introduced at the interaction vertices.}
\label{psich2}
\end{figure}

Finally, Fig.~\ref{psich2}a) shows the 
$J/\Psi{+}N{\to}\Lambda_c{+}\bar{D}$ cross section as a 
function of the invariant collision energy $\sqrt{s}$
calculated with $D$ (solid line) and $D^\ast$ meson 
(dashed line) exchanges and with form factors
at the interaction vertices. The 
$J/\Psi{+}N{\to}\Lambda_c{+}\bar{D}^\ast$ cross section
is shown in Fig.~\ref{psich2}b). Both reactions 
substantially contribute 
at low energies, close to the respective thresholds, while
their contribution to the $J/\Psi{+}N$ dissociation 
cross section decreases with increasing $\sqrt{s}$.  

\section{The $J/\Psi{+}N{\to}N{+}D{+}{\bar D}$ reaction}
The diagram for $J/\Psi$ dissociation by a nucleon with 
the production of the $D{+}{\bar D}$ state is shown 
in Fig.~\ref{psich13}d).
The dissociation involves the $J/\Psi{\to}D{+}{\bar D}$
vertex and therefore it is an OZI allowed process. The total
mass of the produced particles is larger
than the total mass of the initial 
$J/\Psi$-meson and nucleon and thus these reactions
are endothermic. Furthermore, the processes with 
$D{+}N$ and ${\bar D}{+}N$ scattering
are different and both contribute to  
$J/\Psi$ dissociation on the nucleon.

The amplitude for  
the reaction $J/\Psi{+}N{\to}N{+}D{+}{\bar D}$ can be 
parametrized as~\cite{Sibirtsev1,Berestetsky}
\begin{equation}
{\cal M}_d=2g_{JDD}\, \frac{(\epsilon_J 
\cdot p_D)}{t-m_D^2}\,F(t)\, {\cal M_{DN}},
\label{ampf}
\end{equation}
where $p_D$ and $m_D$ are the $D$-meson four-momentum
and  mass, respectively, $\epsilon_J$
denotes the polarization vector of $J/\Psi$-meson, 
$t$ is squared four-momentum
transfered from initial  $J/\Psi$ to final $D$-meson,
while $g_{JDD}$ and $F(t)$ are the coupling constant
and the form factor at the $J/\Psi{D}{\bar D}$ vertex,
respectively.
Furthermore, ${\cal M_{DN}}$ is the amplitude for 
$D{+}N$ or ${\bar D}{+}N$ scattering, which is related
to the physical cross section as
\begin{equation}
|{\cal M_{DN}}|^2 = 16 \pi \,s_1 \,\sigma_{DN}(s_1),
\label{ampf1}
\end{equation}
where $s_1$ is the squared invariant mass of the 
$D{+}N$ or ${\bar D}{+}N$ subsystem.

The double differential cross section for 
the reaction $J/\Psi{+}N\to N{+}D{+}{\bar D}$ 
can be written  as
\begin{eqnarray}
\frac{d^2\sigma_{JN}}{dt ds_1} = \frac{g_{JDD}^2}{96\, \pi^2\,
q_J^2 \, s}\, \,q_D \sqrt{s_1}  \, \frac{F^2(t)}
{(t-m_D)^2} 
\nonumber \\
\times \frac{[(m_J+m_D)^2-t][(m_J-m_D)^2-t]}{m_J^2}\,
\sigma_{DN}(s_1),
\label{crossd}
\end{eqnarray} 
where $q_D$ is given by 
\begin{equation}
q_D^2=\frac{[(m_N+m_D)^2-s_1][(m_N-m_D)^2-s_1]}{4 \, s_1},
\label{on}
\end{equation}
and $m_J$ and $m_N$ denote the $J/\Psi$-meson  and nucleon masses,
respectively.

As was proposed in Ref.~\cite{Berestetsky},
by replacing the elastic scattering cross 
section, $\sigma_{DN}$, in Eq.~(\ref{crossd})
with the total cross section, it is possible to account simultaneously
for all available final states that can be produced at 
the relevant vertex. Since  we are 
interested in inclusive $J/\Psi$ dissociation on the nucleon,
in the following we use the total $D{+}N$ and  ${\bar D}{+}N$ interaction
cross sections. However, we denote this
inclusive dissociation generically as the $J/\Psi{+}N\to N{+}D{+}{\bar D}$ 
reaction.

The total $D{+}N$ and  ${\bar D}{+}N$ cross sections were
evaluated in our previous study~\cite{Sibirtsev2} by
considering the quark diagrammatic approach. 
It was proposed that the $D{+}N$ and  ${\bar D}{+}N$
cross sections should be equal to the ${\bar K}{+}N$ and  $K{+}N$  
cross sections, respectively,  at the same invariant collision energy and
neglecting the contribution from the baryonic
resonances coupled to the ${\bar K}{+}N$ system. The total
${\bar D}{+}N$ cross section was taken as a constant, 
$\sigma_{{\bar D}N}$=20~mb, while the total $D{+}N$
cross section can be parameterized as
\begin{eqnarray}
\sigma_{{\bar D}N} (s_1)
&=&\left(\frac{[(m_{\Lambda_c}+m_\pi)^2-s_1][(m_{\Lambda_c}-m_\pi)-s_1]}
{[(m_D+m_N)^2-s_1][(m_D-m_N)-s_1]}\right)^{1/2} \nonumber \\
&\times & \frac{27}{s_1}+20,
\end{eqnarray}
where $m_L$ and $m_\pi$ are the $\Lambda_c$-hyperon  and
pion masses, respectively, given in GeV and the cross section
is given in mb. 

Some of the partial  $D{+}N$ cross sections were
calculated Ref.~\cite{Ko1} by an effective Lagrangian
approach, where the identity between the  ${\bar K}{+}N$
and $D{+}N$ interaction was imposed by the $SU(4)$ 
relations. Indeed, on the basis of 
$SU(4)$ symmetry the couplings for the vertices
containing ${\bar K}$ or $K$ mesons were taken to be 
equal to those obtained by replacing $\bar{K}$ or $K$ 
with $D$ or ${\bar D}$, respectively. For instance,
$g_{\pi D D^\ast}$=$g_{\pi K K^\ast}$, 
$g_{DN\Lambda_c}$=$g_{KN\Lambda}$, etc. In the more general
case, the large variety of possible diagrams describing
the  ${\bar K}{+}N$ and  $K{+}N$  interactions can be
identified to those for the $D{+}N$ and  ${\bar D}{+}N$
interactions, respectively.  
Thus, the estimates given in Ref.~\cite{Sibirtsev2} are 
supported by the boson exchange formalism.
The comparison between the $DN$ scattering cross section
calculated by an effective Lagrangian
and evaluated by the quark diagrammatic approaches 
is given in Ref.~\cite{Sibirtsevqnp}.  
Furthermore, for the $J/\Psi{+}N{\to}N{+}D{+}{\bar D}$ 
calculations we used coupling constants and 
form factors similar to those used for the
$J/\Psi{+}N{\to}\Lambda_c{+}{\bar D}$ reaction.

\begin{figure}[h]
\phantom{aa}\vspace{-10mm}\hspace{-6mm}
\psfig{file=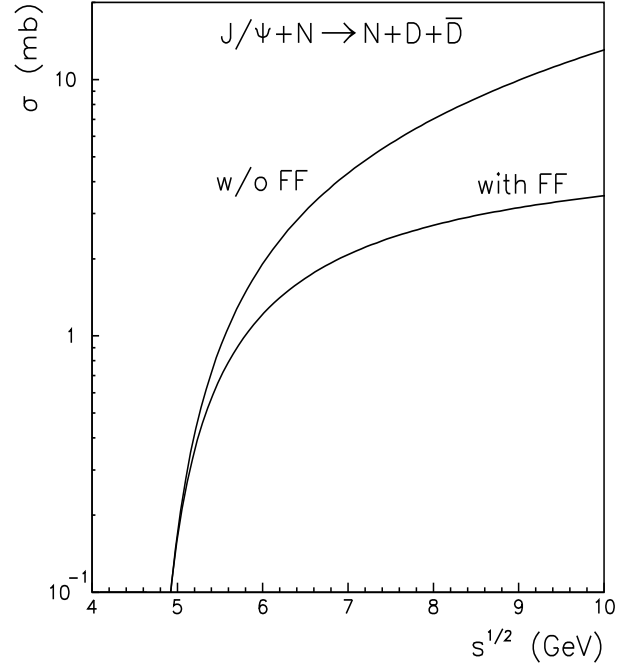,height=10cm,width=9.2cm}
\caption[]{The cross section for $J/\Psi$ dissociation by 
the nucleon with ${\bar D}{+}D$ production as a function
of invariant collision energy $\sqrt{s}$. The results are 
shown for calculations with and without form factor at the
$J/\Psi{\bar D}D$ vertex.}
\label{psich1}
\vspace{3mm}
\end{figure}

Figure~\ref{psich1} shows the cross section for $J/\Psi$ dissociation on 
the nucleon with the production of ${\bar D}{+}D$ pairs calculated using 
Eq.~(\ref{crossd}), with and without form a factor at the
$J/\Psi{\bar D}D$ vertex. The lines in Fig.~\ref{psich1}
show the results where the processes with 
${\bar D}{+}N$ and $D{+}N$ interactions were summed
incoherently. 

\section{Total $J/\Psi{+}N$ cross section and comparison with 
QCD calculations}
\begin{figure}[h]
\phantom{aa}\vspace{-11mm}\hspace{-4mm}
\psfig{file=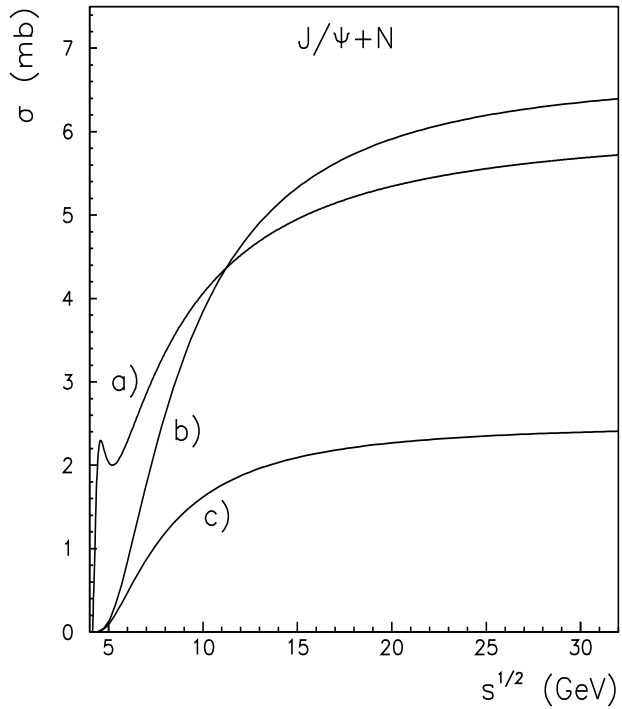,height=10.6cm,width=9.5cm}
\caption[]{The total cross section for $J/\Psi$ dissociation by
the nucleon, as a function of invariant collision energy, $\sqrt{s}$. 
The line a) shows our calculation with a form factors. The line
b) indicates the first order calculations based on a short distance QCD,
while the line c) shows the parameterization from 
Ref.~\protect\cite{Kharzeev1}.}
\label{psich4}
\vspace{3mm}
\end{figure}

Our results for the total $J/\Psi{+}N$ cross section 
are shown in Fig.~\ref{psich4} as the solid line~a). 
Here the total cross section is given as
a sum of the  partial $J/\Psi{+}N{\to}\Lambda_c{+}{\bar D}$ 
cross section calculated with $D$ and $D^\ast$-meson exchange, 
the $J/\Psi{+}N{\to}\Lambda_c{+}{\bar D^\ast}$ cross section
calculated with  $D$-meson exchange and
the inclusive cross section for  $J/\Psi$ dissociation with 
${\bar D}{+}D$ production. 

We found that the total
$J/\Psi{+}N$ cross section approaches  5.5~mb
at high invariant collision energy. We also indicate a partial
enhancement of the $J/\Psi{+}N$ dissociation cross section
at low energies due to the contribution from the
$J/\Psi{+}N{\to}\Lambda_c{+}{\bar D}$ and 
$J/\Psi{+}N{\to}\Lambda_c{+}{\bar D^\ast}$ reaction channels.

The line~c) in Fig.~\ref{psich4} shows the parameterization from
Ref.~\cite{Kharzeev1}, explicitly given as
\begin{equation}
\sigma_{JN}=2.5 \left(1-\frac{\lambda}{\lambda_0}\right)^{6.5},
\end{equation}
where the cross section is given in mb, 
$\lambda_0{=}m_N{+}\epsilon_0$ with $\epsilon_0$ being the
Rydberg energy  and
\begin{equation}
\lambda = \frac{s-m_J^2-m_N^2}{2m_J}. 
\label{para1}
\end{equation}
Parameterization~(\ref{para1}) provides a substantially smaller
$J/\Psi$ dissociation cross section compared to our result.

\begin{figure}[h]
\phantom{aa}\vspace{-11mm}\hspace{-4mm}
\psfig{file=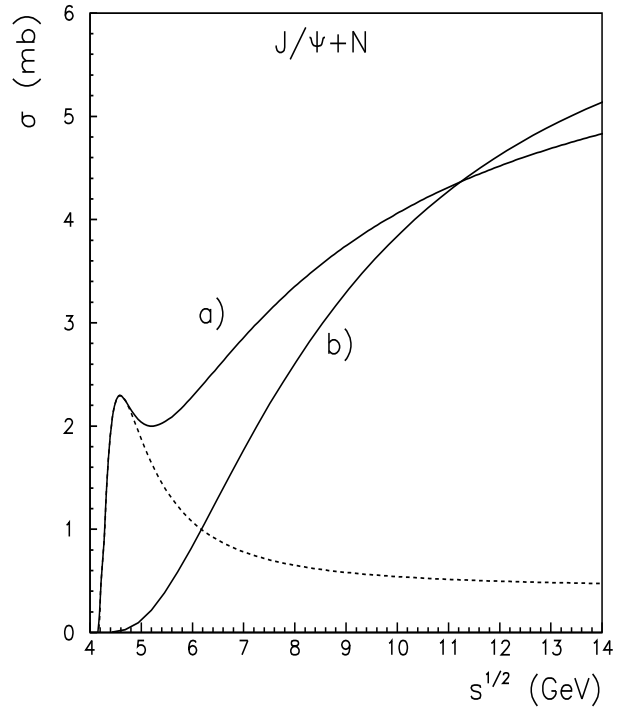,height=10.5cm,width=9.2cm}
\caption[]{The total cross section for $J/\Psi$ dissociation by
the nucleon  as a function of invariant collision energy $\sqrt{s}$. 
The solid line a) shows our calculations with form factors,
while the dashed line indicates the contribution from the 
$J/\Psi{+}N{\to}\Lambda_c{+}{\bar D}$ and 
$J/\Psi{+}N{\to}\Lambda_c{+}{\bar D^\ast}$ reaction channels only. 
The solid line b) indicates the first order calculation 
based on short distance QCD.}
\label{psich4a}
\vspace{3mm}
\end{figure}

Furthermore, the  $J/\Psi{+}N$ cross section can be 
evaluated~\cite{Volkovitsky,Kharzeev2} using short distance QCD
methods based on the operator product expansion~\cite{Peskin}.
Within the first order calculation the cross section
is given as~\cite{Bhanot}
\begin{equation}
\sigma_{JN}=\frac{2^{13}\,\pi}{3^4\,\alpha_s\,m_c^2}\,
\intop^1_{1/\xi}\frac{(\xi x-1)^{3/2}}{(\xi x)^5}
\,\frac{g(x)}{x}\, dx,
\label{para2}
\end{equation}
where $\xi{=}\lambda/\epsilon_0$, $m_c$ is the $c$ quark mass,
$\alpha_s$ is the strong coupling constant and $g(x)$ denotes the
gluon distribution function, for which we
take the form
\begin{equation}
g(x)=2.5(1-x)^4.
\label{pdf}
\end{equation}

Calculations with a more realistic gluon
distribution function~\cite{Martin}  only change
the $J/\Psi{+}N$ cross section slightly at invariant collision
energies $\sqrt{s}{<}$20~GeV, as compared to that obtained 
with function~(\ref{pdf}). The differences associated with various 
gluon structure functions can be predominantly observed
at high $\sqrt{s}$. The $J/\Psi{+}N$ cross section
from the first order calculations performed using short distance
QCD is shown by the line b) in Fig.~\ref{psich4}.

We note that the QCD results are in good agreement with our 
hadronic model calculations at high invariant collision energies,
but substantially deviate from our predictions near the
threshold for endothermic reaction. We ascribe this
discrepancy to the contribution from the  
\newline $J/\Psi{+}N{\to}\Lambda_c{+}{\bar D}$ and
$J/\Psi{+}N{\to}\Lambda_c{+}{\bar D^\ast}$
reactions. The dashed line in
Fig.~\ref{psich4a} shows the separate contribution to the 
total $J/\Psi{+}N$ cross section from the 
$J/\Psi{+}N{\to}\Lambda_c{+}{\bar D}$ and
$J/\Psi{+}N{\to}\Lambda_c{+}{\bar D^\ast}$
reactions, while the solid line a)
again shows our result for the total  $J/\Psi$ dissociation
by a nucleon.
The solid line b) in Fig.~\ref{psich4a} indicates the QCD result
given by Eq.~(\ref{para2}). It is clear that the main
difference between our prediction and the QCD calculation
comes from the $J/\Psi{+}N{\to}\Lambda_c{+}{\bar D}$ 
and $J/\Psi{+}N{\to}\Lambda_c{+}{\bar D^\ast}$ reaction
channels, which contributes substantially at 
low $\sqrt{s}$.

\section{$J/\Psi$ photoproduction on the nucleon}
Within the vector dominance model,  the $J/\Psi$ 
photoproduction invariant amplitude, ${\cal M}_{\gamma{J}}$, on a 
nucleon can be related to the $J/\Psi{+}N$
scattering amplitude, ${\cal M}_{JN}$, as: 
\begin{equation}
{\cal M}_{\gamma J} (s,t) = \frac{\sqrt{\pi \, \alpha}}{\gamma_J} \,
F(t) \, {\cal M}_{JN}(s,t),
\label{vdm}
\end{equation} 
where $s$ is the squared, invariant collision energy, $t$ is
the squared four-momentum transfer, $\alpha$ is the fine-structure constant 
and $\gamma_J$ is the  constant for $J/\Psi$ coupling 
to the photon. In Eq.~(\ref{vdm}) $F(t)$ stands for the
form factor at the $\gamma{-}J/\Psi$ vertex, which accounts for the 
$c{\bar c}$ fluctuation of the photon~\cite{Hufner}. 
Within the naive VDM only the hadronic, but not the
quark-antiquark,  fluctuations of the
photon  through the $\gamma$ mixing with the
vector mesons are considered~\cite{Bauer,Kondratyuk} and the form 
factor $F(t)$ in Eq.~\ref{vdm}
is therefore neglected. As was  calculated in Ref.\cite{Hufner} 
by both the hadronic and quark representations
of the $c{\bar c}$ fluctuation of the photon, the 
form factor $F{=}$0.3 at $t{=}0$. 

The invariant amplitudes are normalized so that the
$J/\Psi{+}N$ differential cross section is written as
\begin{equation}
\frac{d\sigma}{dt}=\frac{|{\cal M}|^2}
{16\,  \pi \,  [(s-m_N^2-m_J^2)^2-4m_N^2m_J^2]},
\end{equation}
and similarly for the $J/\Psi$ photoproduction cross section.

Taking the relation of Eq.~(\ref{vdm}) at $t{=}0$ and
applying the optical theorem for the imaginary part
of the $J/\Psi{+}N$ scattering amplitude, 
one can express the total cross section for $J/\Psi$ dissociation 
by a nucleon $\sigma_{JN}$ in terms of the cross section for
$J/\Psi$ photoproduction on the nucleon at $t{=}0$ as   
\begin{eqnarray}
\left.\frac{d\sigma_{\gamma{N}{\to}JN}}{dt}\right|_{t=0}
=\frac{\alpha}{16\gamma_J^2}\,(1+\alpha_{JN}^2)\,
\, \sigma_{JN}^2\, F^2(0) \nonumber \\
\times \frac{(s-m_N^2-m_J^2)^2-4m_N^2m_J^2}{(s-m_N^2)^2},
\label{photo}
\end{eqnarray}
where $\alpha_{JN}$ is the ratio of the real to imaginary
part of the $J/\Psi{+}N$ scattering amplitude at $t{=}0$. 

The $\gamma_J$ coupling constant can be directly determined 
from the $J/\Psi$ meson decay into leptons~\cite{Nambu}
\begin{equation}
\Gamma (J/\Psi{\to}l^+l^-)= \frac{\pi\, \alpha^2}{3\, \gamma_J^2}
\, \sqrt{m_J^2-4m_l^2}\, \left[1+\frac{2m_l^2}{m_J^2}\right].
\end{equation}

\begin{figure}[h]
\phantom{aa}\vspace{-10mm}\hspace{-6mm}
\psfig{file=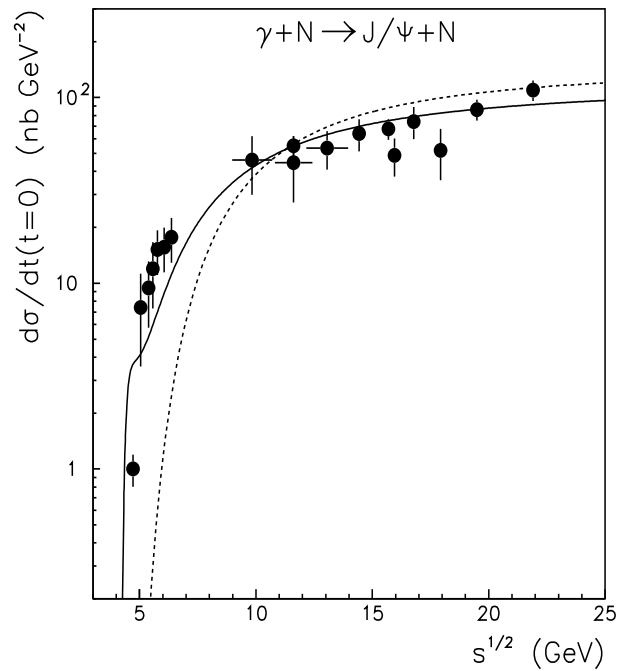,height=10cm,width=9.2cm}
\caption{Differential cross section  for  the reaction 
$\gamma + N \to J/\Psi + N$ at $t{=}0$ as a function of the 
invariant collision energy, $\sqrt{s}$. 
Experimental data are from 
Ref.~\protect\cite{Camerini,Gittelman,Binkley}.
The solid line shows our calculations with the
total $J/\Psi{+}N$ dissociation cross section evaluated using a 
boson exchange model, while the dashed line indicates 
the results obtained for the  $J/\Psi{+}N$ cross section given by
short distance QCD.}
\label{psich5}
\vspace{4mm}
\end{figure}

The ratio $\alpha_{JN}$ is unknown and as a first approximation
we put it equal to zero. This approximation is supported from 
two sides. 

At high energies Regge theory dictates that 
the amplitude for hadron scattering on a nucleon is 
dominated by pomeron exchange and is therefore purely imaginary. 
This expectation is strongly 
supported by the experimental data~\cite{Landshoff} available
for  $p$, ${\bar p}$ $\pi$, $K$ and ${\bar K}$ scattering
by a nucleon. 

At low energies the real part of the $J/\Psi{+}N$  
scattering amplitude at $t{=}0$ can be estimated from the theoretical
calculations of the  $J/\Psi$-meson mass shift  in nuclear 
matter. As was predicted by the calculations by the
operator product expansion~\cite{Luke,Teramond}, QCD van 
der Waals  potential~\cite{Brodsky}
and QCD sum rules~\cite{Klingl,Hayashigaki,Kim}, the mass
of the $J/\Psi$ should only be changed a tiny amount in 
nuclear matter. However, a partial deviation of our results 
from the $ J/\Psi$ photoproduction data might be actually 
addressed to the contribution from  a non vanishing 
ratio $\alpha_{JN}$. More detailed discussion 
concerning an evaluation of the real part of the 
$J/\Psi{+}N$ scattering amplitude will be given in the
following.

The $ J/\Psi$ photoproduction cross section at $t{=}0$ is shown in
Fig.~\ref{psich5} as a function of the invariant collision 
energy. The experimental data were taken from 
Refs.~\cite{Camerini,Gittelman,Binkley}. The solid line
shows the results calculated with the total  $J/\Psi{+}N$
cross section given by the boson exchange model.
We found that within the experimental uncertainties we can 
reproduce the $J/\Psi$ photoproduction data at $t{=}0$ at low
energies quite well. Let us recall that this is possible 
because of the contribution from the 
$J/\Psi{+}N{\to}\Lambda_c{+}{\bar D}$ and 
$J/\Psi{+}N{\to}\Lambda_c{+}{\bar D^\ast}$ reaction channels.

The dashed line in Fig.~\ref{psich5} indicates the calculations 
using Eq.~(\ref{para2}), which reasonably describe the data on
photoproduction cross section at $\sqrt{s}{>}$10~GeV.
We notice a reasonable agreement between our 
calculations and the results from short distance QCD
at invariant collision energies $\sqrt{s}{>}$12~GeV.
We also recall that in Ref.~\cite{Kharzeev2} 
the discrepancy between the QCD results
and the photoproduction data at low $\sqrt{s}$ was 
attributed to a large 
ratio $\alpha_{JN}$. Our result does not support this
suggestion. 

\section{Comparison with Regge theory}
It is of some interest to compare our results with the
predictions from Regge theory~\cite{Donnachie}
given at high energies. Furthermore, for illustrative
purpose we demonstrate the comparison in terms of the
cross section  at $t{=}0$  for the  
$\gamma{+}N{\to}J/\Psi{+}N$ reaction, since the Regge 
model parameters were originally fitted to the
photoproduction data. 

Taking into account the contribution from soft and hard pomeron 
exchanges alone, the exclusive   $J/\Psi$ photoproduction 
cross section  at $t{=}0$ can be explicitly 
written as~\cite{Donnachie1}
\begin{equation}
\left. \frac{d\sigma}{dt}\right|_{t=0}=
23.15s^{0.16}+0.034s^{0.88}+1.49s^{0.52},
\label{pomer}
\end{equation} 
where the cross section is given in nb and the squared 
invariant collision energy, $s$, is in GeV$^2$.
The first term of Eq.~(\ref{pomer}) comes from the soft pomeron
contribution, the second one is due to the hard pomeron, while 
the last term stems from their interference. The parameters 
for the both pomeron 
trajectories were taken from the most recent fit~\cite{Donnachie2}
to the $J/\Psi$ photoproduction data from H1~\cite{H1} and
ZEUS~\cite{ZEUS1,ZEUS2} experiments. 

\begin{figure}[h]
\phantom{aa}\vspace{-10mm}\hspace{-4mm}
\psfig{file=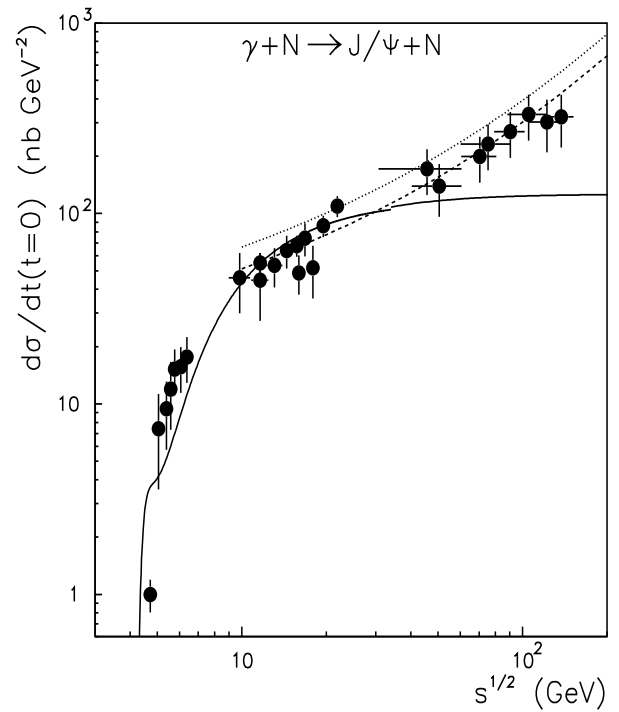,height=10.4cm,width=9.2cm}
\caption[]{The cross section for the reaction \, 
$\gamma{+}N{\to}J/\Psi{+}N$ at $t{=}0$ as a function of 
invariant collision energy, $\sqrt{s}$. 
Experimental data are from 
Ref.~\protect\cite{Camerini,Gittelman,Binkley,Aid,Breitweg}.
The solid line shows our calculations with the
total $J/\Psi{+}N$ dissociation cross section evaluated by
boson exchange model. The dotted line shows the
prediction~(\ref{pomer}) from Regge theory with parameters for the
soft and hard pomeron from Ref.~\protect\cite{Donnachie2},
while the dashed line is the result given by
Eq.~\ref{pomer}, but renormalized by the factor 1.3}
\label{psich6}
\vspace{3mm}
\end{figure}

The prediction from Regge theory is shown by the dotted line
in Fig.~\ref{psich6} together with experimental 
data~\cite{Camerini,Gittelman,Binkley,Aid,Breitweg} on   
$J/\Psi$ photoproduction cross section  at $t{=}0$ as a function 
of the invariant collision energy $\sqrt{s}$. 
The solid line in  Fig.~\ref{psich6} shows our calculations
using the boson exchange model with form factors.
We notice, that in Refs.~\cite{Donnachie,Donnachie2} the Regge model 
fit to the data on $\gamma{+}N{\to}J/\Psi{+}N$ cross section  at $t{=}0$
was not explicitly included in an evaluation of the 
coefficients of Eq.~\ref{pomer}, which might 
partially explain some systematic deviation of the
Regge calculations from the photoproduction data. 

In order to illustrate
the compatibility of the Regge model prediction with 
our calculations  we indicate
by the dashed line in Fig.~\ref{psich6} the result
from Eq.~(\ref{pomer}), renormalized by a factor 1.3.
We note a reasonable agreement between our calculations 
and the Regge theory within the
short range of energies from $\sqrt{s}{\simeq}$10 up to 30~GeV. 

However, considering the result given in this 
section one should keep in mind the following critical 
arguments. 

First, we performed the calculations using a boson 
exchange model, which has some restrictions in its
application at high energies. For instance, the form factors 
introduced at the interaction vertices suppress the contribution 
at large 4-momentum transfer $t$, which allows one to avoid the
divergence of the total cross section at large 
collision energies where the range of the available
$t$ becomes extremely large~\cite{Sibirtsev1}. A more accurate
way to resolve such a divergence at high energies is to use 
the Reggeized boson exchange 
model~\cite{Kaidalov,Sibirtsev4,Kondratyuk2}.

Second, the Regge theory has limitations for applications
at low energies~\cite{Landshoff}. Here we adopt the
parameters for soft and hard pomeron trajectories
that were originally fixed~\cite{Donnachie2} using the large set of
data at $\sqrt{s}{\ge}$40~GeV and the very reasonable agreement between
the Regge model calculations with the data, even at lower
energies, might in some sense be viewed as surprising. 

On the other hand, the $D$ and $D^\ast$ meson exchanges used in
our calculations cannot be related to pomeron 
exchange, rather they can be related to the Regge trajectories, whose
contribution to $J/\Psi$ photoproduction at high 
energies was assumed~\cite{Donnachie,Donnachie1,Donnachie2}
to be negligible. This does not contradict
the results shown in Fig.~\ref{psich6}, since our calculations
are substantially below the data at 
high energies. 
 
\section{Evaluation of the real part of $J/\Psi{+}N$
scattering amplitude}

One of the crucial ways to test the coherence of our calculations 
consists of an evaluation of the real part of the $J/\Psi{+}N$ 
scattering amplitude at $t{=}0$ and a comparison to the various model 
predictions~\cite{Klingl,Hayashigaki,Kim}
for $\Re f(0)$ at zero $J/\Psi$ momentum. The latter is of course 
related to the  $J/\Psi$-meson mass shift or the real part of its
potential in nuclear matter. 

Within the low density theorem the $J/\Psi$ meson mass shift 
$\Delta{m_J}$ in nuclear matter at baryon density  $\rho_B$ can be 
related to the real part of the  scattering amplitude $\Re f(0)$ 
as~\cite{Hayashigaki,Lenz,Dover}
\begin{equation}
\Delta m_J = -2 \pi \frac{m_N+m_J}{m_N \, m_J} \rho_B \Re f(0).
\label{trho}
\end{equation}

The recent QCD sum rule analysis~\cite{Klingl,Hayashigaki,Kim} 
with the operator product expansion
predicts attractive mass shifts of about --4$\div$--10~MeV
for $J/\Psi$ meson in nuclear matter, which corresponds to
small  $\Re f(0)$  about 0.1$\div$0.2~fm. 
This result is very close to the predictions from the operator 
product expansion~\cite{Luke,Teramond} and the calculations with
QCD van der Waals  potential~\cite{Brodsky}. 

There are two ways to link these results for the effective
$J/\Psi$ mass in nuclear matter or the $J/\Psi$-nucleon scattering
length with our calculations. Both of the methods 
applied below contain a number of uncertainties and should be
carefully considered.

First, we address the partial discrepancy between our
calculations of the  $J/\Psi$ photoproduction cross 
section at $t{=}0$ and data at low energies due to the nonzero
ratio $\alpha_{JN}$ appearing in Eq.~(\ref{photo}) 
and evaluate it from the experimental measurements as
\begin{equation}
 \alpha_{JN}^2= \left.\frac{d\sigma^{exp}}{dt}\right|_{t=0}
\times \left[\left.\frac{d\sigma^{th}}{dt}\right|_{t=0}
\right]^{-1}-1,
\label{eval1}
\end{equation}
where the indices $exp$ and $th$ denote the experimental and 
theoretical results for the photoproduction cross section 
at $t{=}0$, respectively.

\begin{figure}[h]
\phantom{aa}\vspace{-10mm}\hspace{-4mm}
\psfig{file=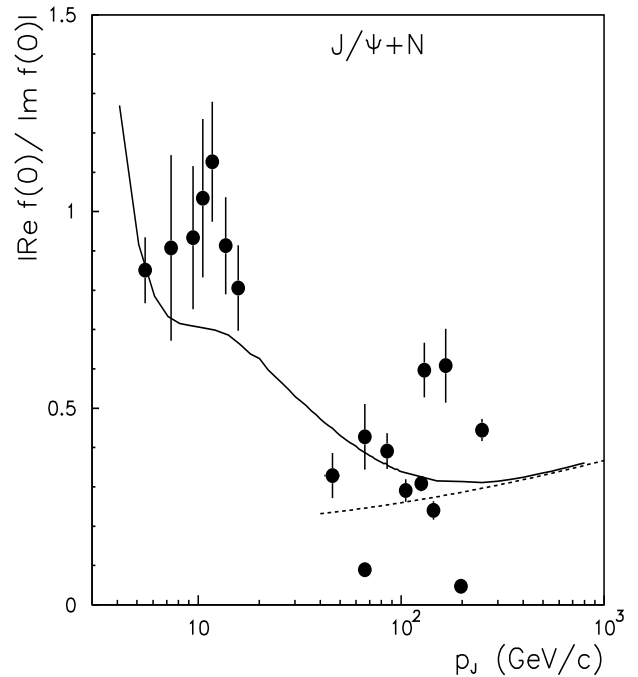,height=10cm,width=9.2cm}
\caption[]{The modulus of the ratio
of real to imaginary part of the  $J/\Psi{+}N$
scattering amplitude at $t{=}0$ extracted from experimental data on  
cross section for  $\gamma{+}N{\to}J/\Psi{+}N$ reaction  as a 
function of $J/\Psi$ momentum $p_J$ in the nucleon rest frame. 
Experimental data used in the evaluation are from 
Ref.~\protect\cite{Camerini,Gittelman,Binkley}.
The solid line shows the results from the  
dispersion relation.
The dotted line shows the prediction from Regge theory,
which fixed the positive sign of the ratio.}
\label{psich8}
\vspace{3mm}
\end{figure}

The  modulus of the ratio $\alpha_{JN}$ evaluated
from the data is shown in Fig.~\ref{psich8} as a function 
of the $J/\Psi$ momentum, $p_J$, in the nucleon rest frame. 
It is clear that the sign of the ratio cannot be fixed 
by Eq.~(\ref{eval1}), but some insight  
might be gained from the Regge theory
calculation, which is shown in Fig.~\ref{psich8}
by the dashed line. Here again we use the 
parameters of the pomeron trajectories from Ref.~\cite{Donnachie2}. 
Now one can conclude that the $\Re f(0)$ is positive at low
$J/\Psi$ momenta, which agrees with the 
predictions~\cite{Luke,Teramond,Brodsky,Klingl,Hayashigaki,Kim} on
the reduction of the  $J/\Psi$ meson mass in nuclear matter.
However, this speculation is true only if the real part of the
$J/\Psi$-nucleon scattering amplitude at $t{=}0$ does not
change its sign at some moderate momentum.

Now, $\Re f(0)$ is given as a product of
the ratio $\alpha_{JN}$ extracted from the photoproduction
data~\cite{Camerini,Gittelman,Binkley} and the imaginary part 
of the $J/\Psi{+}N$
scattering amplitude at $t{=}0$, which is related to the total 
$J/\Psi{+}N$ cross section by the optical theorem~(\ref{optic})
as
\begin{equation}
\Im f(0) = \frac{p_J}{4\, \pi}\,  \sigma_{JN},
\label{optic}
\end{equation}
Let us recall that in our calculations the total
$J/\Psi{+}N$ cross section is given only for OZI allowed 
processes, i.e. at energies $\sqrt{s}{\ge}m_{\Lambda_c}{+}m_D$.
This threshold corresponds to the $J/\Psi$ meson momentum 
of$\simeq$1.88~GeV/c. Thus, within the present method
we cannot  provide the real part of the 
scattering amplitude  at $p_J{=}0$ and 
make a direct  comparison with the absolute value
of the $\Re f(0)$  predicted in 
Refs.~\cite{Luke,Teramond,Brodsky,Klingl,Hayashigaki,Kim}. 
However, the estimate of $\Re f(0)$ at minimal
$J/\Psi$ momenta allowed by OZI reactions might be done.

\begin{figure}[h]
\phantom{aa}\vspace{-10mm}\hspace{-4mm}
\psfig{file=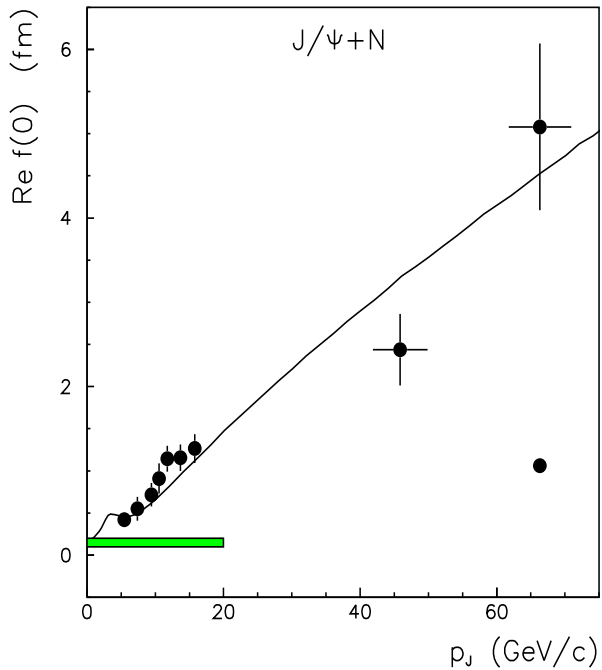,height=10cm,width=9.2cm}
\caption[]{The real  part $\Re f(0)$ of the  $J/\Psi{+}N$
scattering amplitude extracted from experimental 
data~\protect\cite{Camerini,Gittelman,Binkley} on the  
cross section for  $\gamma{+}N{\to}J/\Psi{+}N$ reaction  at $t{=}0$ 
as a function of $J/\Psi$ momentum, $p_J$, in the nucleon rest frame. 
The solid line shows the $\Re f(0)$ evaluated using the
dispersion relation.  The
shadowed area indicates the range of the predictions from 
Refs.~\cite{Luke,Teramond,Brodsky,Klingl,Hayashigaki,Kim} given
at zero momentum of the $J/\Psi$ meson. The extension of the
shadowed area to higher momenta is shown only for orientation
and has no physical meaning.}
\label{psich9}
\vspace{3mm}
\end{figure}

The $\Re f(0)$ extracted from the experimental data on
$J/\Psi$ meson
photoproduction cross section on a nucleon at $t{=}0$ is shown
in Fig.~\ref{psich9} as a function of $J/\Psi$ momentum
in the nucleon rest frame. The shadowed area shows the 
predictions~\cite{Luke,Teramond,Brodsky,Klingl,Hayashigaki,Kim}
at $p_J{=}0$. Notice, that the shadowed area is extended to
higher momenta only for orientation. Within the uncertainties 
of the method we could not detect any discrepancy
with the absolute value of the $\Re f(0)$ given
in Refs.~\cite{Luke,Teramond,Brodsky,Klingl,Hayashigaki,Kim}.
However, we emphasize again, that this conclusion should
be accepted very carefully.

A different way to evaluate the real part of the $J/\Psi{+}N$
scattering amplitude at $t{=}0$ involves the  dispersion
relation that is given as~\cite{Bjorken}
\begin{eqnarray}
\Re f(\omega ) = Re f(\omega_0) +\frac{2(\omega^2-\omega_0^2)}
{\pi} \nonumber \\ 
\times P\intop_{\omega_{min}}^{\infty}
\frac{\omega^\prime \, \Im f(\omega^\prime)}
{({\omega^\prime}^2-\omega_0^2)({\omega^\prime}^2-\omega^2)}
\, d\omega^\prime,
\label{dispa}
\end{eqnarray}
where $\omega$ stands for the $J/\Psi$ total energy,
$\omega_{min}$ is the threshold energy and $\Im f(\omega)$
denotes the imaginary part of the  $J/\Psi$-nucleon
scattering amplitude at $t{=}0$, as a function of energy $\omega$,
which can be evaluated from Eq.~(\ref{optic}). Furthermore,
$Re f(\omega_0)$ is the subtraction constant taken
at $J/\Psi$ energy $\omega_0$.

Being more fundamental, this method involves substantial
uncertainties for the following reasons. First, we should 
specify the subtraction constant. By taking it 
at $p_J{=}0$ from Eq.~(\ref{trho}) we could not further verify 
the consistency of our calculations with the predictions from 
Refs.~\cite{Luke,Teramond,Brodsky,Klingl,Hayashigaki,Kim}.
Second, to evaluate the principal value of the
integral of Eq.~\ref{dispa} one should know the
imaginary part of the  $J/\Psi{+}N$ scattering 
amplitude at $t{=}0$ or the total cross section up to infinite energy.

Since our model could not provide the high energy behavior
of $\sigma_{JN}$ one should make an extrapolation to high
energies by using either 
short distance QCD or Regge theory. As was shown in 
Ref.~\cite{Kharzeev2} the total  $J/\Psi{+}N$ cross section
at high energies is dependent on the gluon
distribution function, $g(x)$, appearing in Eq.~(\ref{para2}).
Furthermore, both distributions given by Eq.~(\ref{pdf}) 
and taken from Ref.~\cite{Martin} were unable to
reproduce~\cite{Kharzeev2} the experimental data on the
$\gamma{+}N{\to}J/\Psi{+}N$ cross section  at 
$\sqrt{s}{>}$70~GeV. Thus we could not extrapolate our 
results to  high energies by the short distance QCD
calculations.

As shown in Fig.~\ref{psich6}, the Regge theory 
reproduces the $J/\Psi$ photoproduction data at $t{=}0$ quite well,
when renormalized by a factor of 1.3. This meets our
criteria since it also fits our calculations at 
$\sqrt{s}{\simeq}$20~GeV. Thus we adopt the Regge model fit
for the extrapolation of our results to high energies in order 
to evaluate the dispersion relation. It is important to notice, 
that, in principle, there is no need to address 
$J/\Psi$ photoproduction in order to extrapolate our results
to infinite energy using Regge theory. One can adopt the 
energy dependence of the dissociation cross section given by 
soft and hard pomeron exchanges and make the absolute
normalization by our calculations $\sqrt{s}{>}$20~GeV.
This is reasonable, because the absolute normalization 
from the Regge model can be considered to some extent as a 
free parameter, not   fixed by the theory.

Moreover, we take the
subtraction constant $\Re f(\omega_0)$  at high energy, 
$\omega_0$=10$^3$~GeV, and fix it by Regge theory
with the predicted ratio $\alpha_{JN}$. Thus we can 
independently use an evaluated  real part of the
$J/\Psi{+}N$  scattering amplitude at $t{=}0$ for comparison 
with the predictions~\cite{Luke,Teramond,Brodsky,Klingl,Hayashigaki,Kim} 
given at $p_J$=0.

Our results for the ratio $\alpha_{JN}$ evaluated by the
dispersion relation~(\ref{dispa}) are shown in
Fig.~\ref{psich8} by the solid line. Obviously, the calculations 
match the Regge model predictions since they were used
to fixed both the high energy behavior of the total
$J/\Psi{+}N$ cross section and the subtraction constant for
the real part of the  scattering amplitude at $t{=}0$.
Furthermore, the dispersion calculations give a surprisingly
good fit to the ratios $\alpha_{JN}$ extracted from the experimental
data~\cite{Camerini,Gittelman,Binkley} on the 
cross section at $t{=}0$ of the $\gamma{+}N{\to}J/\Psi{+}N$ reaction.  

The real part of the  $J/\Psi{+}N$ scattering amplitude
calculated by dispersion relation is shown by the solid
line in Fig.~\ref{psich9} as a function of $J/\Psi$ meson
momentum in the nucleon rest frame. Our calculations 
describes reasonably well the data evaluated from the
$J/\Psi$ photoproduction at $t{=}0$, that are shown 
by the solid circles in Fig.~\ref{psich9}. It is important
that our results, normalized by a subtraction constant at 
high energies, $p_J{=}10^3$~GeV, simultaneously
match the predictions~\cite{Luke,Teramond,Brodsky,Klingl,Hayashigaki,Kim} 
given at $p_J$=0.

Concluding this section let us make a few remarks. 
The two methods applied to the evaluation of the real part of the
$J/\Psi$-nucleon scattering amplitude at $t{=}0$ are 
actually different, but nevertheless provide almost 
identical results for the $\Re f(0)$ over a large range of 
$J/\Psi$ momenta.
The consistency of our approach can be proved by agreement
between our calculations simultaneously with the
Regge model predictions for $\Re f(0)$ at high energy
and the predictions from various 
models~\cite{Luke,Teramond,Brodsky,Klingl,Hayashigaki,Kim}
for $\Re f(0)$ at $p_J{=}0$. 
>From these results we confirm that below the 
momentum of $p_J{\simeq}$1.88~GeV,
which corresponds to the OZI allowed 
$J/\Psi{+}N{\to}\Lambda_c{+}\bar{D}$ threshold, 
the $J/\Psi$ interactions with a nucleon are 
purely elastic. The $J/\Psi$ meson does not undergo significant 
dissociation on absorption at momenta $p_J < 1.88$ GeV.

\section{Estimate of the elastic $J/\Psi{+}N$  cross section}
By definition, the total elastic $J/\Psi{+}N{\to}J/\Psi{+}N$ 
cross section, $\sigma_{el}$, is related to
the elastic differential cross section as 
\begin{equation}
\sigma_{el}{=} \int \left. \frac{d\sigma}{dt}
\right|_{t=0} \exp({bt}),
\label{elact0}
\end{equation}
where $b$ is an exponential slope of the differential
cross section and the cross section at the optical point, $t{=}$0, 
is related to the total $J/\Psi{+}N$ cross section and 
the ratio $\alpha_{JN}$
of the real and imaginary part of the scattering amplitude as
\begin{equation}
\left. \frac{d\sigma}{dt} \right|_{t=0} =
\frac {1}{16 \pi}\, (1+\alpha_{JN}^2) 
\, \sigma_{JN}^2.
\label{elact01}
\end{equation}
The slope $b$ can be estimated from  the differential
$\gamma{+}N\to J/\Psi{+}N$ cross section and was fitted 
in Ref.~\cite{Kharzeev2} to available experimental data as
$b{=}{-}1.64{+}0.83\,\ln s$. 

\begin{figure}[h]
\phantom{aa}\vspace{-10mm}\hspace{-4mm}
\psfig{file=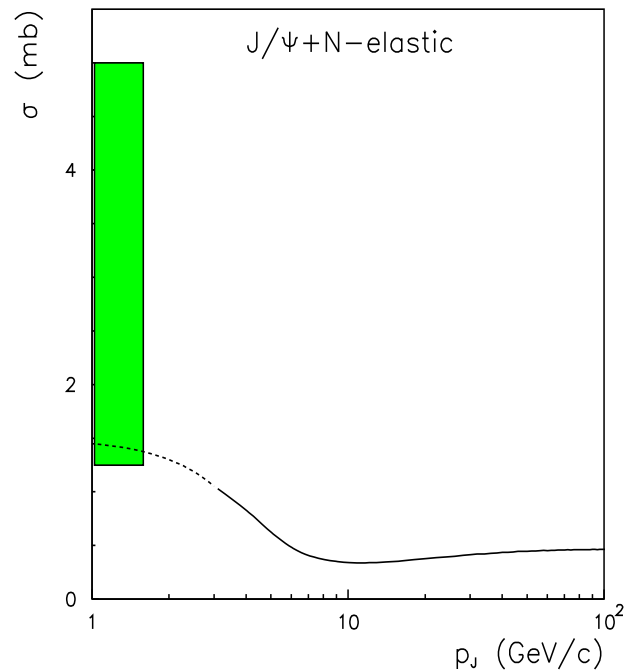,height=10cm,width=9.2cm}
\caption[]{The elastic $J/\Psi{+}N$ cross
section as a function of the $J/\Psi$ meson momentum in the 
nucleon rest frame. The solid lines show our calculations.
The dashed area illustrates predictions from 
Refs.~\protect\cite{Luke,Teramond,Brodsky,Klingl,Hayashigaki,Kim} given
at $p_J{=}0$. The dashed line indicates our interpolation for
the elastic cross section based on the QCD sum rule 
results alone~\protect\cite{Klingl,Hayashigaki,Kim}.}
\label{psich11a}
\vspace{2mm}
\end{figure}

Finally, the elastic $J/\Psi{+}N$ 
cross section is given by
\begin{equation}
\sigma_{el}{=} \frac {1}{16 \pi\, b}\, (1+\alpha_{JN}^2) 
\, \sigma_{JN}^2,
\label{elact1}
\end{equation}
and is shown in Fig.~\ref{psich11a} as a function of $J/\Psi$ meson
momentum. The dashed area in Fig.~\ref{psich11a} indicates the
elastic cross section at $p_J{=}0$ given within the scattering 
length approximation in  
Refs.~\protect\cite{Luke,Teramond,Brodsky,Klingl,Hayashigaki,Kim} as
\begin{equation}
\sigma_{el}=4 \pi |\Re f(0)|^2,
\label{elact2}
\end{equation}
that is valid as far as $\Im f(0)$=0 at $p_J$=0. Note,
that Eq.~\ref{elact1} could not provide the estimate
for elastic cross section at $p_J{<}$1.88~GeV, i.e.
below the $\Lambda_c{+}\bar{D}$ reaction threshold,
where the dissociation cross section $\sigma_{JN}$=0. 
In principle, one can interpolate
$\sigma_{el}$ from zero $J/\Psi$ momentum to our results.

As we already discussed, the uncertainty in the elastic cross
section  at $p_J$=0, given in 
Refs.~\cite{Luke,Teramond,Brodsky,Klingl,Hayashigaki,Kim}, 
is very large and to make a less unambiguous 
interpolation let us to use only the most recent results from
the QCD sum rule calculations~\cite{Klingl,Hayashigaki,Kim}.
Let us recall that the QCD sum rule calculations from 
Ref.~\cite{Klingl} indicate a shift of the 
$J/\Psi$ mass of about --5$\div$--10~MeV, while in 
Ref.~\cite{Hayashigaki} the mass shift was
reported to be in the range --$7{\le}\Delta{m_J}{\le}$--4 MeV.
The calculations with higher dimension operators in the QCD sum rule
provide~\cite{Kim} $\Delta m_J$=--4~MeV. 
Thus, we take an average value which corresponds to 
$\sigma_{el}$ =1.6~mb at $p_J$=0. Now the dashed line  in  
Fig.~\ref{psich11a} indicates our interpolation of the elastic 
cross section at low momenta.

\begin{figure}[h]
\phantom{aa}\vspace{-11mm}\hspace{-4mm}
\psfig{file=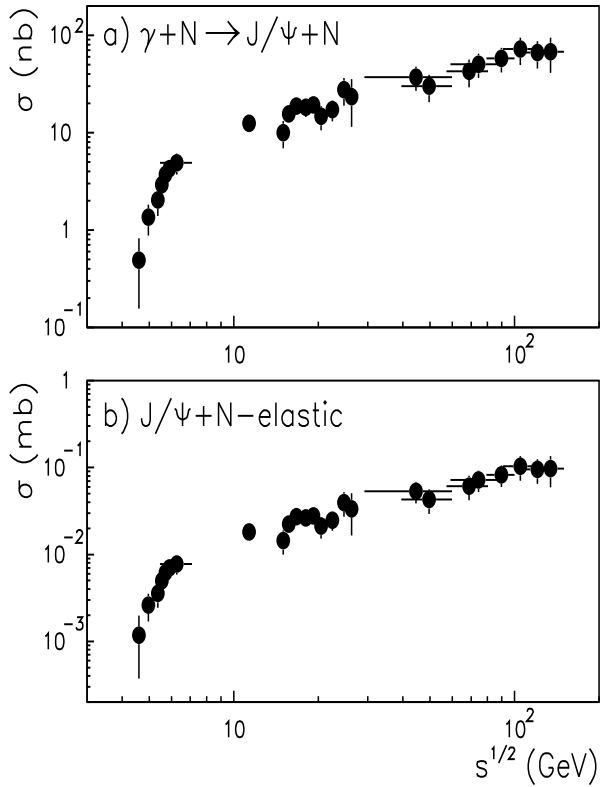,height=11.8cm,width=9.2cm}
\caption[]{The total $\gamma{+}N{\to}J/\Psi{+}N$ cross section
(a) collected in Ref.~\protect\cite{Alde} 
and the elastic $J/\Psi{+}N{\to}J/\Psi{+}N$ 
cross section (b) evaluated by naive vector dominance
given by Eq.~\ref{naive} as
a function of the invariant collision energy $\sqrt{s}$.}
\label{psich10}
\vspace{2mm}
\end{figure}

Furthermore, the vector dominance model 
can be used to relate the elastic $J/\Psi{+}N{\to}J/\Psi{+}N$ 
cross section $\sigma_{el}$ and the total 
$\gamma{+}N{\to}J/\Psi{+}N$ cross section $\sigma_{\gamma{J}}$. 
Applying Eq.~(\ref{vdm}) and neglecting the form
factor this relation can be written as 
\begin{equation}
\sigma_{el}=\frac{\gamma_J^2}{\pi \, \alpha}
\frac{(s-m_J^2)^2}{(s-m_N^2-m_J^2)^2-4m_N^2m_J^2} \,
\sigma_{\gamma{J}}
\label{naive}
\end{equation}

Fig~\ref{psich10}a) shows the $\gamma{+}N{\to}J/\Psi{+}N$ 
cross section~\cite{Alde} as a function of invariant collision energy, 
while the elastic cross section $\sigma_{el}$ evaluated
from experimental photoproduction data using Eq.~(\ref{naive})  
is shown in Fig~\ref{psich10}b). Apart of the absolute value
of the $\sigma_{el}$, which is substantially below our 
estimate, we notice that very strong energy dependence of the
elastic $J/\Psi{+}N{\to}J/\Psi{+}N$ cross section
evaluated using Eq.~(\ref{naive}) might alone indicate the inconsistency
of the application of naive VDM~\cite{Hufner}. Indeed,
as we discussed above, the calculations of the $J/\Psi$ 
mass shift in matter or $J/\Psi{+}N$ scattering length given in 
Refs.~\cite{Luke,Teramond,Brodsky,Klingl,Hayashigaki,Kim} 
predicts the elastic cross section in the range
$1.25{\le}\sigma_{el}{\le}$5~mb at $p_J$=0, while the $J/\Psi$ meson
photoproduction data evaluated by Eq.~(\ref{naive})
indicate a systematic decrease of $\sigma_{el}$ with
decrease of $J/\Psi$ meson momentum. 

\begin{figure}[h]
\phantom{aa}\vspace{-10mm}\hspace{-4mm}
\psfig{file=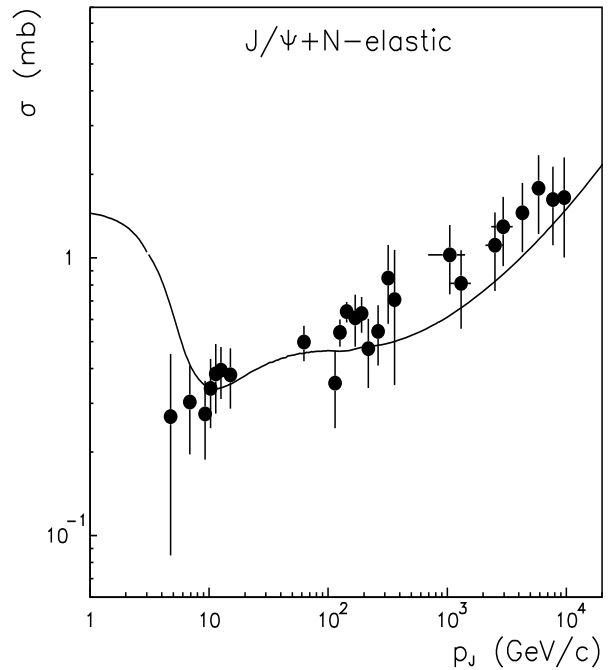,height=10cm,width=9.2cm}
\caption[]{The circles show the elastic $J/\Psi{+}N{\to}J/\Psi{+}N$ 
cross section as a function of the  $J/\Psi$ momentum
evaluated from the total $\gamma{+}N{\to}J/\Psi{+}N$ cross section
by vector dominance model of Eq.~\ref{naive} additionally
corrected by the factor $\kappa$ given by  Eq.~\ref{correct}. 
The solid line indicates our estimate for $\sigma_{el}$
calculated by Eq.~\ref{elact1}.}
\label{psich12}
\vspace{3mm}
\end{figure}

Apparently this inconsistency might be resolved by 
introducing a form factor $F(t)$ at the $\gamma{-}J/\Psi$ 
vertex~\cite{Hufner}, as given by Eq.~(\ref{vdm}) and 
applied in our study. The  correction  to the naive vector 
dominance is then given as
\begin{equation}
\kappa{=} \left[ \int \exp(b \, t) \right]^{-1} \,
\int F^2(t)\exp\, (b\, t),
\label{correct}
\end{equation}
where $b$ is the slope of the differential $\gamma{+}N{\to}J/\Psi{+}N$ 
cross section and  the integration in Eq.~(\ref{correct}) 
is performed over the range of the squared transverse momentum $t$
available at given invariant collision energy $\sqrt{s}$.

By taking, for simplicity, an exponential form of the 
$\gamma{-}J/\Psi$  form factor with cutoff parameter
$\Lambda$=1.7~GeV$^{-2}$, we evaluate the elastic 
$J/\Psi{+}N{\to}J/\Psi{+}N$ cross section from 
the data on the $\gamma{+}N{\to}J/\Psi{+}N$ 
cross section~\cite{Alde} and show the result by
the circles in Fig.~\ref{psich12}, as a function
of the $J/\Psi$ meson momentum. The solid line 
in Fig.~\ref{psich12} indicates our estimate
for $\sigma_{el}$, which clearly give a reasonable fit to 
the correctly extracted 
data over a large range of available momenta $p_J$. 

It is worthwhile to note that Fig.~\ref{psich12} clearly
illustrates that the introduction of a form factor
at the $J/\Psi{-}N$ vertex allows one to resolve an  
inconsistency in the evaluation the elastic $J/\Psi{+}N{\to}J/\Psi{+}N$ 
cross section from the  total $\gamma{+}N{\to}J/\Psi{+}N$ 
cross section.


\section{Final results}

\begin{figure}[h]
\phantom{aa}\vspace{-11mm}\hspace{-4mm}
\psfig{file=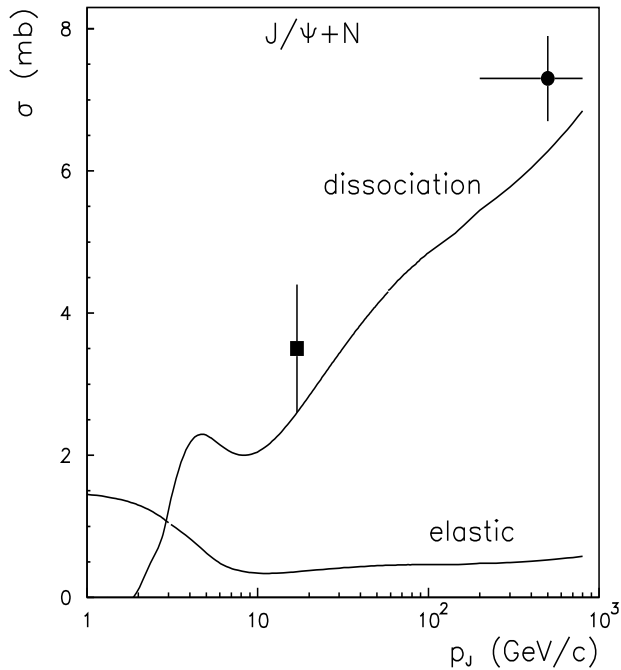,height=10cm,width=9.2cm}
\caption[]{The elastic and dissociation $J/\Psi{+}N$ cross
sections as a function of the $J/\Psi$ meson momentum in the 
nucleon rest frame. The solid lines show our results.
The square shows the $J/\Psi{+}N$ dissociation 
cross section evaluated from $J/\Psi$ photoproduction on nuclei,
while the circle indicates the dissociation cross section 
evaluated from $J/\Psi$ production from proton-nucleus
collisions.}
\label{psich11}
\vspace{3mm}
\end{figure}

Finally,   Fig.~\ref{psich11} shows the  
$J/\Psi{+}N$ dissociation and elastic cross sections 
over a large range of $J/\Psi$ momenta.
Let us recall that the dissociation 
cross section, $\sigma_{JN}$, was calculated using the
boson exchange model up to $p_J{\simeq}200$~Gev/c and extrapolated
to high energies by Regge theory. 

Now we  compare our results with the
$J/\Psi{+}N$ dissociation cross section evaluated from 
nuclear reactions. The square in Fig.~\ref{psich11} shows
the $\sigma_{JN}$ extracted from the $J/\Psi$ photoproduction 
from nuclei at a mean photon energy of 17~GeV~\cite{Anderson}.
The $J/\Psi$ meson momentum was not explicitly fixed
and is indicated in  Fig.~\ref{psich11} only for orientation.

Furthermore, the circle  indicates the result from the combined 
analysis~\cite{Kharzeev4} of experimental data on $J/\Psi$ 
production from $p{+}A$ collisions at beam energies 
from 200 to 800 GeV. Again the $J/\Psi$ meson momentum
was not explicitly fixed by the analysis~\cite{Kharzeev4}.

Within the experimental uncertainties our 
results for the total  $J/\Psi{+}N$  dissociation cross section
fit the available experimental data reasonably well. 

We notice, that our results slightly ($\simeq$15\%) underestimate 
the $J/\Psi{+}N$  dissociation cross section evaluated from 
experimental data on $J/\Psi$ production from
$\gamma{+}A$ and $p{+}A$ reactions.

We also note that the  NA50 experimental data on $J/\Psi$ 
production from heavy ion collisions~\cite{NA50} can be 
reasonably well fitted by the  
calculations~\cite{Cassing,Capella,Vogt}
with a $J/\Psi{+}N$ dissociation cross section
$\sigma_{JN}{\simeq}3{\div}6.7$~mb. The most recent
calculations~\cite{Cassing1,Capella1,Sibirtsev9}
on $J/\Psi$ production and comparison to 
NA50 data~\cite{NA50} indicates a smaller dissociation cross
section, in the range from 3 to 4.5~mb. In central $Pb{+}Pb$
collisions at 160 A$\cdot$GeV the $J/\Psi$ meson momenta $p_J$ in the 
nucleon rest frame  are distributed over the range from
15 to 70~GeV/c with the maximum at
$p_J{\simeq}$40~GeV/c~\cite{Cassingp}. Comparing the heavy ion
results~\cite{NA50,Cassing1,Capella1,Sibirtsev9,Cassing,Capella,Vogt} 
with our calculations from Fig.~\ref{psich10} we  
also find reasonable agreement with $A{+}A$ data.

Furthermore, the nonperturbative QCD calculations~\cite{Navarra1}
provides the $J/\Psi{+}N$ dissociation cross section of
$4.4{\pm}0.6$ mb at $J/\Psi$ meson momenta above 200~GeV,
which is consistent with our results shown by Fig.~\ref{psich11}.

\section{Future perspectives}
There are a few problems that remain open and 
need further investigation. 

In the present study we considered  only OZI allowed processes resulting
in an endothermic inelastic reaction, where the total mass of the 
produced particles is larger than the initial mass, $m_J{+}m_N$. 
These reactions provide the threshold behavior of the cross
section. On the other hand, the OZI suppressed reaction channels such as
\newline $J/\Psi{+}N\to N+ n\pi$ with the number of the final
pions from $n$=1 up to $m_J{/}m_\pi{\simeq}22$  are open
even at $p_J$=0.  These exothermic reactions, where the
total final mass of the produced particles is less that the
initial mass, provide a $1/p_J$ behavior of the $J/\Psi{+}N$ 
dissociation cross section. 

For instance, the OZI suppressed 
$J/\Psi{+}N{\to}N{+}\pi$ cross 
section was calculated 
by the $\rho$ meson exchange model~\cite{Sibirtsev8}. 
It was found that the
cross section is negligibly small although it diverges at
$p_J$ close to zero, because it is exothermic. 
Furthermore, the $\rho$ meson exchange model evaluates
the $J/\Psi{\to}\rho{+}\pi$ vertex, which accounts only  
tiny fraction, $\simeq$1.27\%, for the total hadronic decays of the
$J/\Psi$ meson. The multi-meson OZI suppressed reactions
might play a non negligible role for $J/\Psi$ dissociation
on a nucleon. Eventually additional theoretical estimates 
are necessary in order to make a final conclusion about the
possibility of $J/\Psi{+}N$ dissociation at low momenta.

Furthermore, we found a  strong  momentum dependence of the
$J/\Psi{+}N$ dissociation cross section, as illustrated by 
Fig.~\ref{psich11}. This momentum dependence of $\sigma_{JN}$ 
might provide a partial explanation of the 
variation  of the slope $\alpha$ from the $A^\alpha$-dependence
with the Feynman variable $x_F$. As was 
observed~\cite{Alde} in proton-nucleus collisions at beam 
energy of 800~GeV the extracted value for $\alpha$ is $\simeq$0.9 
at $x_F$=0.2, while it decreases ${\simeq}0.8$ at 
$x_F{\simeq}$0.7. The Feynman variable $x_F$ is proportional to the
$J/\Psi$ meson  momentum and one may naturally expect from 
the results shown in Fig.~\ref{psich11} that at large
momenta, $p_J$, or equivalently large $x_F$, the value for 
$\alpha$ extracted from the $A^\alpha$-dependence, should be smaller in
comparison to small $p_J$. Now we can
qualitatively predict that the slope $\alpha$ decreases
with increasing $x_F$. However, a quantitatively
analysis needs further calculation.

An alternative way is to reanalyze the experimental data
on the $A$-dependence of $J/\Psi$ production from $p{+}A$
collisions and to evaluate the dissociation cross section 
as a function of the $J/\Psi$ momentum. This will provide
a crucial test of our calculations, since it can be
directly compared to results shown in Fig.~\ref{psich11}.

Furthermore, the ratio $\alpha_{JN}$ of the real to imaginary
part of the  $J/\Psi{+}N$ scattering amplitude at $t{=}0$,
which is shown by Fig.~\ref{psich8}, in principle
can be measured using coherent
$J/\Psi$ production at diffractive minima~\cite{Bauer}.    
Coherent $J/\Psi$ photoproduction from nuclei seems 
to be an optimal way for such a measurement, however one needs 
to perform further investigations in order to fix the sensitivity
of the data to the sign and magnitude of the ratio $\alpha_{JN}$.

Obviously, the form factor $F(t)$ at the $\gamma{-}J/\Psi$ vertex
plays a key role for the comparison of our hadronic
calculations with the data on 
$\gamma{+}N{\to}J/\Psi{+}N$ cross section. While this
form factor is theoretically motivated at $t{=}$0 by
the microscopic calculations given in Ref.~\cite{Hufner},
the evaluation  of its $t$-dependence still need
further studies. Our results shown in Fig.~\ref{psich12}
should be consider as a phenomenological estimate 
for $F(t)$.

In addition, one can address the effect due to in-medium modification 
of the $J/\Psi{+}N$ dissociation cross section because of  the
strong changes of the $D$ and $D^\ast$ meson properties
in nuclear matter~\cite{Sibirtsev2,Tsushima9}. As was 
found in Ref.~\cite{Sibirtsev9} this modification 
plays a substantial role for $J/\Psi$ dissociation 
on comovers. One might expect that the modification 
of the $J/\Psi{+}N$ dissociation cross section
in nuclear matter might play no role for heavy ion collisions 
where the available $J/\Psi$ momenta 
given in the nucleon rest frame are larger than  $p_J{>}$13~GeV, 
as was discussed above. However, to clarify the situation
a detailed calculation of the in-medium modification
of the $J/\Psi{+}N$ dissociation amplitude is necessary.

\section{Summary}

We have calculated the $J/\Psi{+}N$ dissociation cross section by
a boson exchange model, including $D$ and $D^\ast$
meson exchange and considering the $\Lambda_c{+}{\bar D}$ 
and $N{+}D{+}{\bar D}$ final states. We note that our results 
are in reasonable agreement with short distance QCD
calculations at high energies, while differing from them
at low energies because of the 
$J/\Psi{+}N{\to}\Lambda_c{+}{\bar D}$ reaction channel
explicitly included in our model.

To compare our results with the data on 
$\gamma{+}N{\to}J/\Psi{+}N$ cross section we introduced 
form factor at the $\gamma{-}J/\Psi$ vertex proposed
in Ref.~\cite{Hufner} and  found good agreement with the
photoproduction data. 

Furthermore, we evaluated the real part of the 
$J/\Psi{+}N$ scattering amplitude at $t{=}0$, $f(0)$, 
using a dispersion relations
and extrapolating our result for the dissociation cross section 
to infinite energy by the Regge theory~\cite{Donnachie,Donnachie1}. 
Fixing the 
subtraction point for $\Re f(0)$ at large energies
with Regge theory, we were able to simultaneously saturate
the $\Re f(0)$ at zero momentum of $J/\Psi$ meson given
by the predictions for the $J/\Psi$ mass shift
in nuclear matter and for the $J/\Psi{+}N$ scattering
length~\cite{Luke,Teramond,Brodsky,Klingl,Hayashigaki,Kim}.

We estimate the elastic $J/\Psi{+}N{\to}J/\Psi{+}N$ cross section 
from our calculations and illustrate the compatibility 
of our results with the data on  the total 
$\gamma{+}N{\to}J/\Psi{+}N$ cross section.
It is important to note that the $J/\Psi$ photoproduction
data were used in our study mostly for illustrative purposes 
and do not influence the parameters of our calculations 
or the final results.

Finally, we predict the energy dependence of the $J/\Psi{+}N$
dissociation, $\sigma_{JN}$, and the elastic cross section 
over a large range of energies, from 1 to 10$^3$~GeV. In contrast
to the usual expectation of a constant dissociation
cross section, we found that $\sigma_{JN}$ varies strongly 
over the indicated range of energies. Our results
are in agreement with the  $\sigma_{JN}$ evaluated 
phenomenologically from 
the experimental data on $J/\Psi$ meson production in
$\gamma{+}A$, $p{+}A$ and $A{+}A$ collisions. 

\acknowledgements
A.S would like to acknowledge the warm hospitality and
partial support of the CSSM during his visit.
The discussions with  E. Bratkovskaya, W. Cassing, 
C. Greiner, D. Kharzeev and J.~Speth are appreciated. 
This work was supported by the Australian Research Council and the 
Forschungzentrum J\"{u}lich.

\end{document}